\documentclass[review]{elsarticle}
\pdfoutput=1
\usepackage{graphicx} 
\usepackage{braket}
\usepackage{caption}
\usepackage{lineno,hyperref}
\usepackage{physics}
\usepackage{float}
\usepackage{times}
\usepackage{listings}
\usepackage[margin=3cm]{geometry}
\usepackage{qcircuit}
\usepackage{natbib}
\usepackage{algorithm}
\usepackage{algpseudocode}
\usepackage[T1]{fontenc}
\usepackage{gensymb}
\usepackage[caption=false,font=normalsize,labelfont=sf,textfont=sf]{subfig}
\usepackage{qcircuit}
\usepackage{graphicx}
\usepackage{tabularx,booktabs,textcomp}

\hypersetup{pdfauthor={Name}}

\journal{Quantum Information Processing}
\begin{document}
\begin{frontmatter}
\title{A Novel Approach to Threshold Quantum Images by using Unsharp Measurements}
\author[1,3]{Ayan Barui}
\author[2]{Mayukha Pal}
\author[3]{ Prasanta K Panigrahi\corref{mycorrespondingauthor}}
\ead{pprasanta@iiserkol.ac.in}
\cortext[mycorrespondingauthor]{Corresponding author}
\address[1]{Central University of Haryana, SH 17, Jaat, Haryana 123031, India}
\address[2]{ABB Ability Innovation Center, Asea Brown Boveri Company, Hyderabad 500084, India}
\address[3]{
Indian Institute of Science Education and Research Kolkata, Mohanpur 741246, India}


\begin{abstract}
    {We propose a hybrid quantum approach to threshold and binarize a grayscale image through unsharp measurements (UM) relying on image histogram. Generally, the histograms are characterized by multiple overlapping normal distributions corresponding to objects, or image features with small but significant overlaps, making it challenging to establish suitable thresholds. The proposed methodology uses peaks of the overlapping Gaussians and the distance between neighboring local minima as the variance, based on which the UM parameters are chosen, that maps the normal distribution into a localized delta function. To demonstrate its efficacy, subsequent implementation is done on noisy quantum environments in Qiskit. This process is iteratively repeated for a multimodal histogram to obtain more thresholds, which are then applied to various life-like pictures to get high-contrast images, resulting in comparable peak signal-to-noise ratio and structural similarity index measure values. The obtained thresholds are used to binarize a grayscale image by using novel enhanced quantum image representation integrated with a threshold encoder and an efficient quantum comparator (QC) that depicts the whole binarized picture. This approach significantly reduces the complexity of the proposed QC and of the whole algorithm when compared to earlier models.}
\end{abstract}

\end{frontmatter}




\section{Introduction}\label{sec1}

Efficient image processing has received significant attention in recent times due to the proliferation of images in different environments. To extract important image features and characteristics, we require image segmentation that divides an image into a smaller set of pixels, emphasizing its vital image characteristics while reducing storage size \cite{SAHOO1988233}. Extensive research has focused on image binarization and multi-thresholding techniques of digital images to represent widely distributed pixel values by a few numbers, which helps in image compression, enhances image contrast, hence reduces the cost of image transmission \cite{wiley2018computer}. This improvement is observed in image features because distributed pixel values around the maxima in the image histogram generally originate from objects and image features. Classification and feature extraction methodologies apply to the image's information if it is invariant to position, orientation, and scale to ensure proper image processing. However, the need to significantly increase the efficiency of image processing and analysis has become particularly pressing given the continuously growing size of picture data and the increasingly difficult computational jobs \cite{ruan2021quantum}.

Quantum image processing (QIMP) has emerged as an essential field, aiming to harness the potential of quantum computers for efficient storage, representation, and processing of digital images, overcoming constraints present in classical computations \cite{le2011flexible,sun2011multi,zhang2013neqr}. Early advancements include Vlasov's work in 1997, which identifies orthogonal images  \cite{Vlasov}. Venegas-Andraca proposed a way to store four different values in a single qubit to realize speedup using QIMP, using the idea of qubit lattice  \cite{article}. The two most widely used image encoding algorithms that serve as the base of QIMP are flexible representation of quantum images (FRQI) and novel enhanced quantum representation of digital images (NEQR). FRQI, proposed by Lee et al. in 2010, uses phase encodings to represent grayscale information in probability amplitudes of qubits  \cite{le2011flexible}. The multi-channel representation for quantum images (MCRQI) \cite{sun2011multi}, an extension to the colorized version of FRQI, was proposed the following year. Although FRQI encodes the pixel information into one qubit, it is difficult to perform an operation on one qubit without changing the states of other qubits. To overcome this problem, NEQR \cite{zhang2013neqr} was proposed in 2013. Although more qubits are used to represent an image, the implementation of NEQR is much simpler than other representations, which is why it is used widely, particularly in image processing tasks. Normalized amplitude encodings of images is demonstrated in the work of Srivastava et al. \cite{srivastava2013quantum}. Hybrid phase-based quantum image representations is also an efficient technique to lower the complexity cost of representation \cite{mandal2023hybrid}. A number of improved FRQI and NEQR models, like the improved novel enhanced representation of quantum images using INCQI, were also developed \cite{su2021improved}. Using these quantum representations, many different algorithms have been proposed to binarize an image. Pixel values must be compared with a particular threshold to determine two or more regions in the image histogram where the pixels are segmented.\\

Wang showcased the practical realization of a quantum comparator's ability to distinguish between two input numbers using Toffoli gates  \cite{wang2012design}. Al Rabadi demonstrated the application of reversible logic using a convolution-based encoder in Quantum Logic Network Implementation of the Viterbi Algorithm \cite{al2009closed}. A more optimized quantum comparator, proposed by Thapliyal, reduced the complexity in the logarithmic scale \cite{5697872}. A comparison of 2 $n$-bit quantum logic states using only a single ancillary bit was demonstrated by Xia et al. \cite{xia2018efficient}. Vudadha showed the idea of using reversible adders using the ripple carry in his design of a quantum comparator, which he also used to make adders \cite{6296477}. These designed quantum comparators could be used to binarize an image, where the pixel and threshold values are compared. Initial works comprise the methods proposed by Simona Caraiman, which is a way to threshold a quantum image using histogram-based segmentation \cite{CARAIMAN201446} and the other being a proposition of a quantum circuit to segment a given image \cite{caraiman2015image}. Many works based on quantum image thresholding made through the NEQR or INEQR framework include the binarization technique by Du et al. \cite{DU2022105710}, who devised a quantum circuit to flip the state of a qubit depending on an ``activation state" which is used to binarize an image. A dual threshold algorithm in which the ancillary qubits do not scale up with increasing image size was shown and demonstrated using simulators by Yuan et al. \cite{yarticle}. A quantum version of the Otsu algorithm was proposed in the work of Panchi Li \cite{li2020design}, where the results are verified using classical simulations. The design of fault-tolerant quantum comparators by optimizing the number of T gates is seen in Orts's work \cite{orts2021optimal}. Recent developments include a moving average method-based adaptive threshold for a quantum image segmentation algorithm, which is simulated on the IBM Quantum Experience platform using the Qiskit extension \cite{yuan2022adaptive}. Other techniques include thresholding using quantum comparators optimized efficiently by the ripple carry idea, as shown by Xia et al. \cite{doi:10.1142/S0217732320500492}. The aforementioned quantum algorithms for binarizing the images use a pre-defined threshold value. In our work, we have defined a threshold value by encoding the pixel values in the image's histogram to quantum states and performing UM over it. We use the probability amplitudes of the resulting quantum state to determine the threshold.\\

The proposed procedure of UM involves a gradual focus on the average pixel value via the interpretation of the probability amplitudes of the quantum states, mirroring how various distributions tend to converge towards the Dirac delta function as given below:
\begin{equation}
\label{delta}
    \underset{\sigma\to 0}{lim} \frac{1}{ \sqrt{(2\pi\sigma^2)}}e^{-\frac{(x-\mu)^2}{2\sigma^2}}= \delta(x-\mu)
\end{equation}

In our approach to image multi-thresholding, a novel hybrid mode has been employed, primarily relying on the analysis of image histograms as it is the most rudimentary form of data available for analysis. Secondly, the methodology presented in this study can be characterized as ``hybrid" as it involves the processing of data to generate histograms. The histograms of typical images are a set of overlapping Gaussian distributions. Hence, using Gaussian positive operator valued measures (POVMs) is ideal for discerning these normal distributions and reducing them to a particular singular value, as shown in Eq.(\ref{delta}). 
These Dirac delta functions are assigned as threshold values and are used to compress the given image. Furthermore, the optimal number of thresholds is inherently determined by the number of Gaussians within the histogram, establishing optimal image segmentation as compared to other processes. The quality of the compressed images is comparable to algorithms like the fast statistic recursive method \cite{ARORA2008119} and multi-Otsu method \cite{6313341} in terms of PSNR and SSIM \cite{1284395}. In the second part of our work, we have proposed a quantum comparator, which is a significantly improved version of  Xia et al.'s \cite{doi:10.1142/S0217732320500492} work. The proposed QC determines which pixel is greater than the obtained threshold, and using the output, we have produced a binarized form of a grayscale image. In representing the binarized picture, we have used only one bit to represent the intensity values of the binarized picture, thus reducing the circuit depth and cost compared to other quantum binarization algorithms. This procedure is implemented in the Qiskit platform, which demonstrates the effectiveness of the proposed quantum comparator.\\

The paper's structure is given as follows: Section \ref{s2} discusses some preliminary concepts and essential topics necessary for comprehending the work. In Section \ref{s3}, the proposed method of computing a threshold from the histogram of a given image is discussed and elaborated with examples. In Section \ref{s4}, we use quantum circuits to binarize an image and implement it on the Qiskit platform with the help of the proposed quantum comparator. Section \ref{s5} discusses the quality of the images using PSNR and SSIM obtained by compressing various life-like and noisy images and comparing them with other well-established classical algorithms. Additionally, the complexity analysis of the whole algorithm along with the proposed quantum comparator is also discussed with their previous counterparts. Section \ref{s6} concludes the work.

\section{Materials and Methods}
\label{s2}
In this section, a few topics are discussed briefly to focus on thresholding and different comparison metrics used in this work and to gain insights into unsharp measurements, particularly formulating the discrete operator basis. 
\subsection{Binarization and Multi-Thresholding}
A dot or square on any display screen represents a pixel. They are the basic building blocks of a digital image or display and are created using geometric coordinates. A pixel has a value, which is the measure of intensity depending on which it emits light. For an 8-bit grayscale image, we have $2^8=256$ levels of intensity. A 0-intensity pixel represents a complete black pixel, and a 255-intensity pixel depicts a fully illuminated pixel. Binarization serves as a technique for representing a grayscale image using only 0 and 255 pixels, reducing the image's size while preserving informational data. As a preliminary step, image preprocessing is employed to enhance performance. Preprocessing is classified into three categories: noise removal, background estimation, and color-to-grayscale conversion. For simplification purposes, we assume that all these three categories are already satisfied, with comments arising while discussing the proposed algorithm. There are many algorithms for the binarization of an image like ``Adaptive Thresholding" by Sauvola \cite{SAUVOLA2000225}, local thresholding by Niblack \cite{niblack1986introduction}, and ``Otsu's Method" \cite{4310076} which are used widely in image processing. Recent developments include block thresholding method by Hemachander et al. \cite{HEMACHANDER2007119}. Applications of this binarization procedure are realized in the edge detection approach (canny edge detection \cite{Wang2009AnIC}), which is used to find the boundary between foreground and background. Recently, Shouvik et al. made the use of cuckoo search method to segment images \cite{chakraborty2023balanced}.\\

\begin{figure}[h]
    \centering
    \includegraphics[scale=0.5]{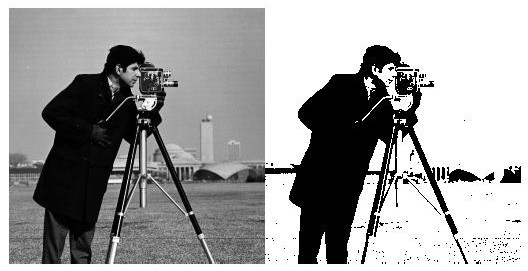}
    \caption{Image of cameraman and its binarized version, which successfully captures the informational content of the original image}
\end{figure}

Another way to efficiently store and represent grayscale images is the technique of multi-level thresholding, which segments a grayscale image into several distinct regions and sets more than two thresholds for a single image. The mentioned procedure works quite well when the picture has a complex background and bi-level thresholding incorrectly or partially distinguishes between the foreground and background pixels. Examples of this algorithm are seen in the works of Reddi et al. \cite{6313341}, where the Otsu method is generalized to multi-level thresholding, and a statistical method is described in the work of Arora et al., which gives good compression results in PSNR in less computational time \cite{ARORA2008119}.
\subsection{Peak Signal-to-Noise Ratio (PSNR) and Structural Similarity Index Measure (SSIM)}
The PSNR measures the quality of a compressed image with respect to its original form. The amount of distortion or error generated during signal processing or transmission is measured objectively using the PSNR ratio. It is a popular metric for image compression. The Mean Squared Error ($x$) between the unprocessed and processed signals determines PSNR which is calculated in Decibels ($dB$) using the following formula:
\begin{equation}
    PSNR = 10\log_{10}(\frac{mp^2}{x}) dB
\end{equation}
 where $mp$ is the highest pixel value ($mp = 255$ for an 8-bit grayscale image). A value of 100 $dB$ suggests that the original and compressed images have no difference. The larger the value, the better the compressor.
 SSIM is a frequently used statistic for assessing the visual similarity of two images, which was introduced in the paper \cite{1284395}. Conventional metrics like PSNR predominantly focus on pixel-wise variations, which may not align with human perception. In contrast, SSIM considers both structural information and brightness variations.\\
\\
The SSIM gives a range of values from -1 to +1, quantifying the similarity between two given images. A score of +1 signifies a near-perfect or identical image, whereas -1 score indicates significant dissimilarity between the two images. These numbers are frequently modified to fall inside the range [0, 1], where the extremes have the same significance. Furthermore, the SSIM methodology employs a window-to-window-based similarity approach that analyzes two images, with windows representing small segments of the images. The factors for comparing these windows are given below in the form of an equation considering two windows $x$ and $y$:\\
\begin{equation}
SSIM(x,y) = \frac{(2\eta_x\eta_y + c_1)(2\lambda_{xy} + c_2)}{(\eta_x^2 + \eta_y^2 + c_1)(\lambda_x^2 + \lambda_y^2 + c_2)}    
\end{equation}
where $\lambda_x$ is the pixel sample mean of x, $\lambda_x^2$ is the variance of x, $\lambda_y$ is the pixel sample mean of y, $\lambda_y^2$ is the variance of y, $\lambda_{xy}$ is the covariance of x and y, and finally, $c_1$ and $c_2$ are two stabilizing variables. An application of SSIM is seen in the classification of chest X-ray images \cite{unknown}.

\subsection{Unsharp Measurements}
\label{section:Feature}

In this study, we will work with quantum states, describing the given system thoroughly. Given a quantum state,  we can decompose it into a discrete set of basis states, which is described by the following equation: 
\begin{equation}
\ket{\psi}=\frac{1}{A_i}\sum_i\ket{x_i}
\end{equation}
where {$\ket{x_i}$} represents the basis states and $A_i$ is the normalizing factor which normalizes the quantum state $\ket{\psi}$. Any projector could be defined on the basis of a given system as $\ket{x_i}\bra{x_i}$. They are represented by $\Pi_i=\ket{x_i}\bra{x_i}$ with the conditions that $\Pi^\dag=\Pi$ and ${\Pi}^2=\Pi$ which when satisfied are called a projector and with the condition that $\sum_i{\ket{x_i}\bra{x_i}}=\sum{\Pi_i}=I$ yields a projective measurement (PVM) which represent sharp measurements on a basis.\\

In standard textbook of quantum mechanics, only sharp/projective measurement is considered; in other words, we measure every reading accurately. However, in practical cases, measurements are inaccurate by nature. Therefore, we often need to work in a more general framework known as quantum unsharp measurement. We define a general quantum unsharp event as a weighted average of quantum sharp events. Positive operator value-based measurements (POVM) are a set of operators that measure the system unsharply and do not cause a complete collapse of the state but change the original quantum state. POVMs discriminate between nonorthogonal basis states, which the PVMs cannot. Suppose $E_m$ are such a set of operators, so they must satisfy the following conditions that $E_m=E_m^\dag$ for all $m$ and $\bra{\psi}E_m\ket{\psi}\geq0$ for all $m$ and all $\ket{\psi}\in \mathcal{H}$. Additionally,  $\sum_m{E_m}=I$ is called a partition of unity. Applications of UM are seen in Wigner's form of the Leggett-Garg inequality \cite{PhysRevA.91.032117}.\\

Fuzziness could be introduced in any system via the use of effect operators, explained in the works of effect algebras, which was first formulated by Ludwig \cite{ludwig1983ensembles} and further developed by Giuntini and Greuling \cite{giuntini1989toward}. They are be defined in the discrete basis $\ket{x_i}$ as:
\begin{equation}
    E_y = \sum_{q_i \in (00...0)}^{(11...1)} \frac{1}{ \sqrt{(2\pi\delta^2)}}e^{-\frac{(y-x_i)^2}{2\delta^2}}\ket{x_i}\bra{x_i}
\end{equation}

The determinants influencing the effect operators are denoted as $\delta$ and $y$. In conventional terms, $\delta$ represents the Gaussian distribution's standard deviation, conceptually understood as the Gaussian's spread or, equivalently, as the degree of fuzziness we intend to incorporate into the system. On the other hand, $y$ embodies the distribution's mean, signifying the specific value of the discrete basis where the Gaussian peak is located.

\section{Computing Threshold of a Digital Image}
\label{s3}

In this section, our primary objective is to explain the approach for determining the threshold values for image segmentation using POVMs applied to constructed states. To comprehensively represent an image, we rely on two key data sources, namely the spatial arrangement of pixels, which helps establish the image's dimensions and the intensity values assigned to these pixels. When establishing the threshold values, we focus solely on the intensity values. To effectively process the information on a quantum computer, it is essential to map the data (information of pixels) into quantum states. The image's histogram is crucial in supplying valuable input to the algorithm, particularly when constructing the Gaussians. The entire procedure of state preparation and measurements could be executed following a methodology similar to using unsharp measurements in clustering data points by Srushti et al. \cite{patil2023measurement}, as follows:
 \begin{itemize}
     \item Let us take any digital $n$-bit grayscale image and encode each intensity value as its binary equivalent into the qubits by making an equal superposition of all the possible states ($2^n$ states) in $\ket{I}$. Each intensity value is encoded as a single quantum state $\ket{Q_m}$, which is the binary equivalent of its corresponding pixel. Then, the sum of all such states represents the complete intensity information of any $n$ -bit grayscale image. Note that to calculate thresholds, we need only the intensity information, not the position information, so we need not take any additional qubits. 
\begin{equation}
\label{e6}
    \sum_{Q_m=00...0}^{11...1}\ket{Q_m}=\ket{I}
\end{equation} 
In Eq. (\ref{e6}), $\ket{I}$ is an unnormalized quantum state containing every possible intensity value between 0 and $2^n - 1$ and $m \in [0,n-1]$.
So we obtain state $\ket{I}$ containing all the binary encodings of selected pixel intensity values in the given image. 
    \item Following the preparation of the state $\ket{I}$, which remains unnormalized, the determination of the necessary qubit count depends on the number of states contained within  $\ket{I}$. The number of qubits required to encode all the values would be $n$. Moving on to the preparation of effects and effect operators, it is worth noting that these operators, as previously discussed, incorporate two critical parameters: the width $\delta$ and the mean $i$ of the distribution, as depicted in Eq. (\ref{eq6}). 

\begin{equation}\label{eq6}
    E_i= \sum_{Q_m \in (00...0)}^{(11...1)} \frac{1}{ \sqrt{(2\pi\delta^2)}}e^{-\frac{(i-m)^2}{2\delta^2}}\ket{Q_m}\bra{Q_m}    
\end{equation}

    \item To determine these two values, the histogram of the provided image is utilized. The mean value of the effect is identified as the intensity level corresponding to a peak in the histogram. This approach is guided by the objective of thresholding the image based on intensity values that exhibit a higher frequency than other pixels, which are easily found by analyzing the image's histogram. The distribution width ($\delta$) is set to an optimal value, depending on the width of a certain overlapping Gaussian. After the preparation of an effect operator, it is applied to the above-prepared states of the given digital image $\ket{I}$, introducing fuzziness near the selected intensity values, which is denoted by $i$ in Eq.(\ref{eq6}) in the discrete basis $x_i$, where $x_i$ is the range of intensity values present in the given image. A similar approach using the peak detection method is seen in M. I. Sezan's work of an automated peak detection algorithm \cite{SEZAN199036} and multi-thresholding by the use of image histogram in the work of Papamarkos \cite{PAPAMARKOS1994357}.

 After applying the effect operators on the state $\ket{I }$, we get a new state as:
 
  \begin{equation}\label{eq7}
    E_i\ket{I}= \sum_{Q_m \in (00...0)}^{(11...1)} \frac{1}{ \sqrt{(2\pi\delta^2)}}e^{-\frac{(i-m)^2}{2\delta^2}}\ket{Q_m}\bra{Q_m}\sum_{m=0}^{n}\ket{Q_m}
 \end{equation}

 \begin{equation}\label{eq8}
     E_i\ket{I}= \frac{A}{ \sqrt{(2\pi\delta^2)}} \sum_{m=0}^{n}e^{-\frac{(i-m)^2}{2\delta^2}}\ket{Q_m}
 \end{equation}

    \item The final expression in Eq. (\ref{eq7}) is still not yet normalized. As we started with an unnormalized state, we added a normalizing factor $A$ in the expression in Eq. (\ref{eq8}).
\end{itemize}

This process is repeated if there is more than one distribution in the histogram. The \textbf{Algorithm \ref{algo1}} is depicted to outline the above process in the form of a pseudocode and explained via two examples in \textbf{Case-1} and \textbf{Case-2}. $p, r$, and $d$ contain information about the various distributions in the histogram, $thresh$ is an empty list that will contain the threshold values depending on the number of peaks. If the number of peaks is one (unimodal histogram), the state with the maximum counts after measurement becomes the threshold value. Multiple $E_i$'s are constructed and operated on $\ket{I}$ for multiple peaks.\\

\textbf{The Divide and Conquer Algorithm} is an alternative approach to get the measurement results. After constructing the state in lines 9 and 16, we devised a different approach to simulate this on a quantum computer because implementing a Gaussian POVM on a quantum circuit is complex. However, works on state preparation by Pinto et al. \cite{PhysRevA.107.022411} have shown one qubit two-element and one qubit three-element implementation of POVM. An effective state preparation based on the ``divide and conquer" idea by Araujo et al. is used in which any quantum state could be represented on a quantum circuit by appropriate usage of rotation and phase gates \cite{araujo2021divide}. The angles are chosen according to the probability amplitudes of the states obtained from applying the POVM on the initial state $\ket{I}$. These amplitudes are used to generate angles to perform operations that transform $\ket{0}_n=\ket{0}^{\otimes n}$ into any $n$-dimensional vector given in Eq. (\ref{phi}) by the use of the gen\_angle algorithm given in Araujo et al.'s work. 
\begin{equation}\label{phi}
    \ket{\phi}=x_0\ket{0}+...+x_{N-1}\ket{N-1}
\end{equation}

\begin{algorithm}
	\caption{Generate threshold values} 
    \label{algo1}
        \textbf{Input:} $n$-bit grayscale image\\
        \textbf{Output:} Threshold values
        
        \begin{algorithmic}[1]
        \State Plot histogram of the given image and obtain $p,r$ and $d$
        \State $p$ is the no. of peaks in the histogram in image histogram
        \State $r$ is a list containing all the intensity values at which the peaks occur
        \State $d$ is a list containing the corresponding widths
        \State Initialize a list $thresh$ 
        \State prepare $n$ qubits in $2^n$ combinations
        \If {$p=1$}
        \State $i=r[0]$ \textbf{and} $\delta=d[0]$
        \State Construct $E_i$ and operate it on $\ket{I}$ as given in Eq.(\ref{eq8})
        \State {\fontfamily{qcr}\selectfont
        gen\_angles($E_i\ket{I}$)
        }
        \State Simulate by divide and conquer load circuit, do measurement
        \State $thresh \gets$ quantum state with maximum counts
        \Else 
		\For {$k \gets 0$ \textbf{to} $p$}
        \State $i=r[k]$ \textbf{and} $\delta=d[k]$
        \State Construct $E_i$ and operate it on $\ket{I}$ as given in Eq.(\ref{eq8})
         \State {\fontfamily{qcr}\selectfont
        gen\_angles($E_i\ket{I}$)
        }
        \State Simulate by divide and conquer load circuit, do measurement
        \State Store results in dictionary $dict_k$
	\For {key$_u$, value$_u$ \textbf{in} $dict_k$}
            \For{key$_v$, value$_v$ \textbf{in} $dict_{k+1}$}
				\If{key$_u$=key$_v$ \textbf{and} value$_u$=value$_v$ }
				\State Store key$_u$ in $thresh$
			\EndIf
		\EndFor
        \EndFor
        \EndFor
        \EndIf
        \end{algorithmic}
        
\end{algorithm}

{\fontfamily{qcr}\selectfont
gen\_angles()
}is used to generate angles by recursively computing the amplitudes of the input vector via the inverse sine function, taking values from the binary leaf tree. We use this algorithm only to simulate the obtained state in lines 9 and 16 of the algorithm. Detailed procedure for the exact calculation of the angles could be found in the reference \cite{araujo2021divide}. To understand the procedures we take two cases, one being an unimodal histogram and the other being a bimodal or multimodal histogram.\\

\begin{figure*}
\centering
\captionsetup[subfigure]{font=small}
\subfloat[]{\includegraphics[height=0.2\textwidth]{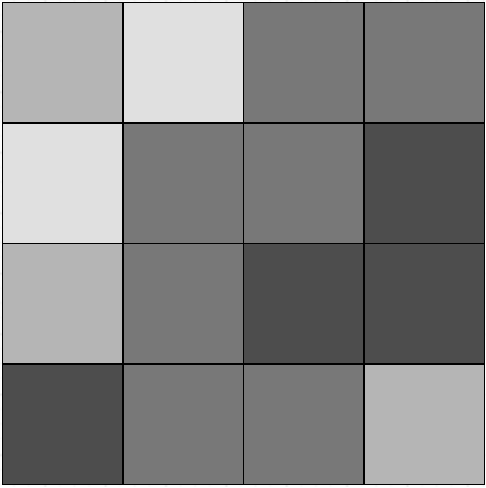} 
\label{f2a}}
\quad
\subfloat[]{\includegraphics[height=0.3\textwidth]{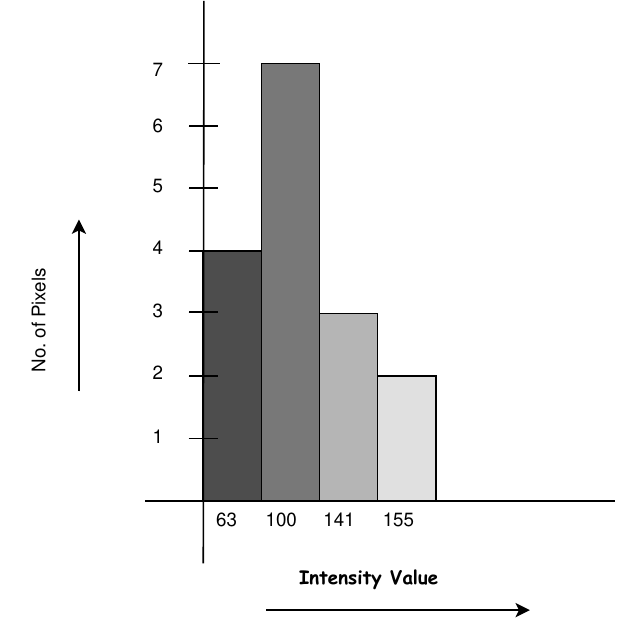} 
\label{f2b}}
\caption{(a) This is a $4 \times 4$ grayscale image with some certain intensities namely 63, 100, 141, 155, (b) The corresponding histogram of the number of pixels and their intensities}
\end{figure*}

\textbf{Case-1}\\
 Let us consider a $4\times 4$ image given in Fig. \ref{f2a} containing 16 pixels with intensities 63, 100, 141, and 155 and their corresponding encodings given in Table \ref{t1}, which represents the intensity present in the grayscale image. Here, the binary equivalence of the intensity value is not taken as it will increase the number of qubits required to encode intensity information. Instead, as there are four values present, we design the encoding scheme to make $n=2$ ($n=log_2(4)$) as small as possible.

 \begin{table}[h]
        \centering
         \caption{Encodings of the intensity values present in Figure \ref{f2a}}
        \begin{tabular}{|c | c |}
        \hline
        Intensity Value & Encoding  \\
        \hline \hline
        63 & $\ket{00}$ \\
        100 & $\ket{01}$ \\
        141 & $\ket{10}$ \\
        155 & $\ket{11}$ \\
        \hline
       \end{tabular}
      
       \label{t1}
    \hfill
\end{table}
In Fig. {\ref{f2b}}, the histogram illustrates the distribution of pixel intensities of the image in Fig. {\ref{f2a}}. As previously mentioned, our objective is to binarize the image around the peak intensity, which, in this case, is 100, representing the most prevalent intensity value in the 4 × 4 picture. Thus, we select the threshold value $i$ as 100, considering all discrete intensity values within the image. It is important to note that any intensity value not listed in Table {\ref{t1}} is absent in the image and is set to 0 using $q_i$. We have chosen $\delta$ as 35 and $A$ as 0.6795 for optimization purposes. Substituting these values into Eq. (\ref{eq8}), we obtain the final state as follows:

\begin{equation}
\label{e12}
    E_i\ket{I} = 0.447\ket{00} + 0.774\ket{01} + 0.387\ket{10} + 0.223\ket{11}
\end{equation}

This state can also be rewritten as:
\begin{equation}\label{eq10}
 \ket{I}=\sqrt{0.2}\ket{00} + \sqrt{0.6}\ket{01} + \sqrt{0.15}\ket{10} + \sqrt{0.05}\ket{11}
\end{equation}

 It is seen that $\ket{01}$ has the highest probability amplitude, and according to the motivation of the algorithm, we choose the corresponding intensity as the threshold. The final prepared state is different from our initial state $\ket{I}$, and it tells us which state is the most probable to be the threshold. This state can be simulated through the divide and conquer algorithm by constructing the leaf tree  

 \begin{figure}[h]
\centering
\captionsetup[subfigure]{font=small}
\subfloat[]{\includegraphics[scale=0.65]{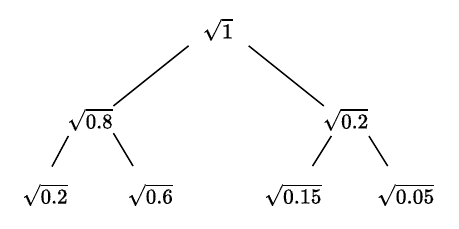}
\label{bl}}
\quad
\subfloat[]{\includegraphics[scale=0.75]{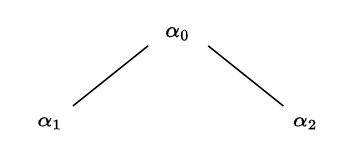} \label{al}}
\caption{(a) Bottoms up recursive calls for generating $\alpha$ angles for the state in Eq. (\ref{eq10}), (b) Corresponding generated angles generated by the function {\fontfamily{qcr}\selectfont
gen\_angles()
}  } 
\end{figure}

From the {\fontfamily{qcr}\selectfont
gen\_angles()
} \ algorithm we have calculated that $\alpha_0=2\sin^{-1}[{\frac{\sqrt{0.05}}{\sqrt{0.15}+\sqrt{0.05}}}]=120\degree$; $\alpha_1=2\sin^{-1}[{\frac{\sqrt{0.6}}{\sqrt{0.2}+\sqrt{0.6}}}]=60\degree$ and $\alpha_2=2\sin^{-1}[{\frac{\sqrt{0.2}}{\sqrt{0.8}+\sqrt{0.2}}}]=53.13\degree$.

\begin{figure}[htbp]
    \centering
    \includegraphics[scale=0.5]{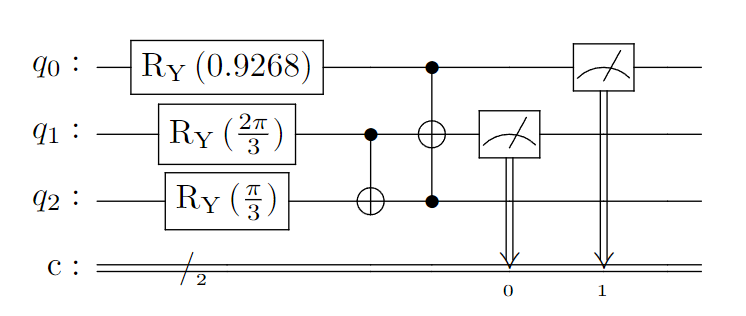}
    \caption{The circuit representation that prepares the state given in Eq.(\ref{eq10}). The corresponding angles that are structured are given as 0.927 rad, $2\pi/3$ rad, and $\pi/3$ rad. The CNOT gates are used to distribute the phases correctly over the leaf tree given in Araujo et al.'s work \cite{araujo2021divide}}
    \label{f4}
\end{figure}

After the angles have been defined, we load these angles onto the divide and conquer load circuit in lines 11 and 18 of algorithm \ref{algo1} to make a quantum circuit by using suitable R$_y$ and CNOT gates. The state given by Eq. (\ref{eq10}) is simulated in Fig. \ref{f4}.\\

\begin{figure}[htbp]
    \centering
    \includegraphics[scale=0.35]{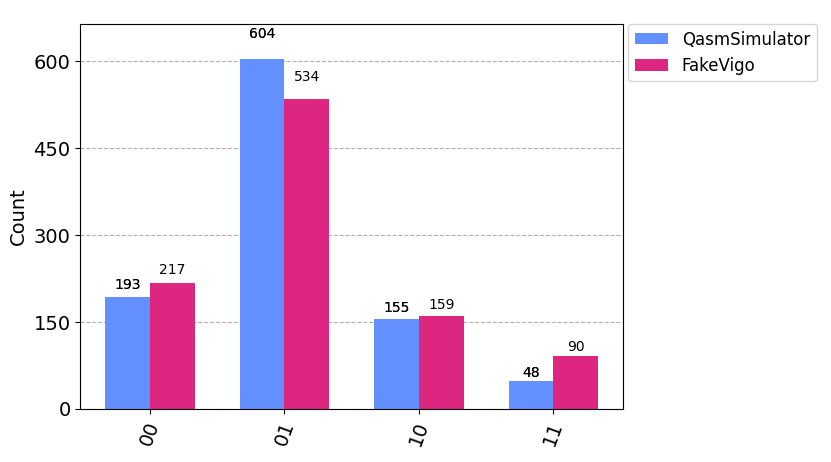}
    \caption{Demonstration of the output of the circuit mentioned in Figure {\ref{f4}}. The state $\ket{01}$ has the highest probability amplitude of 60\% (out of 1000 shots). We obtained this result by simulation using the QasmSimulator present in Qiskit compared with the FakeVigo backend which mimics real noisy quantum computers. The results of both the backends are at par with each other  }
\end{figure}
 
The threshold value for binarization is established by identifying the state with the highest count, which, in this case, corresponds to the state $\ket{01}$ representing the pixel with an intensity of 100. However, it is straightforward to determine the threshold for four-pixel values with an unimodal histogram. The process becomes more intricate for a $256 \times 256$ image, requiring multiple threshold values to segment the image more efficiently as the image histogram would be multimodal in nature.\\

\begin{figure}[h]
\centering
\captionsetup[subfigure]{font=small}
\subfloat[]{\includegraphics[scale=0.40]{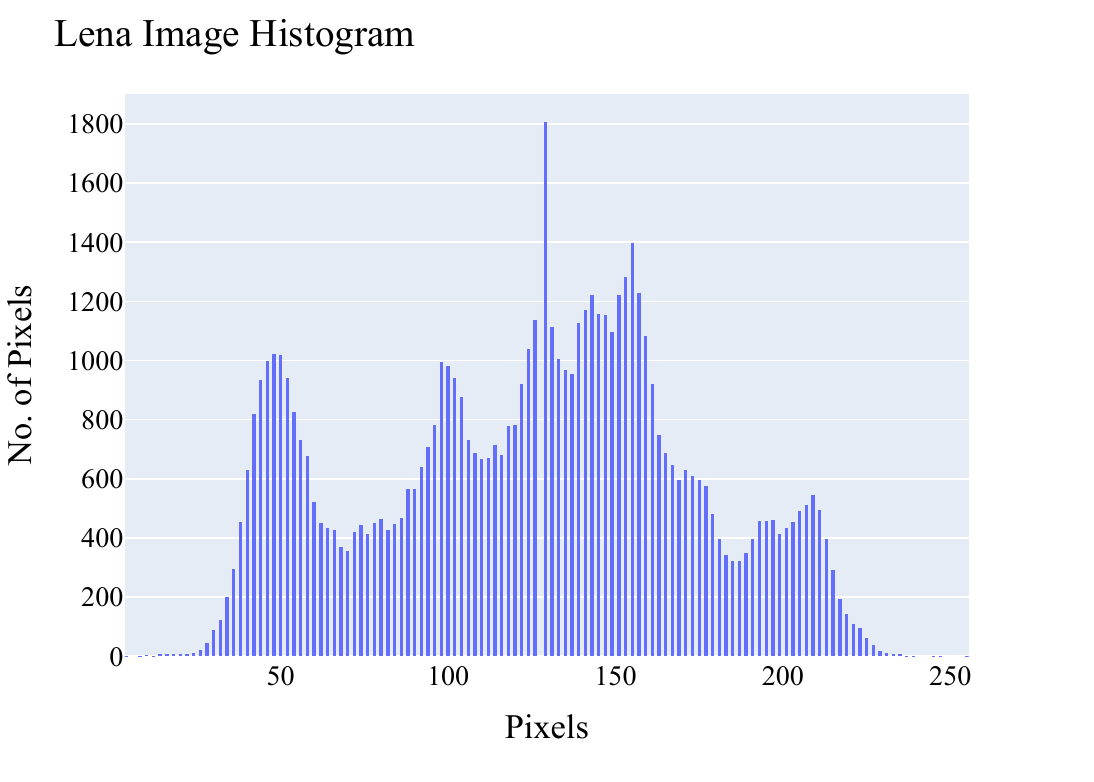}
\label{f6a}}
\quad
\subfloat[]{\includegraphics[scale=0.3]{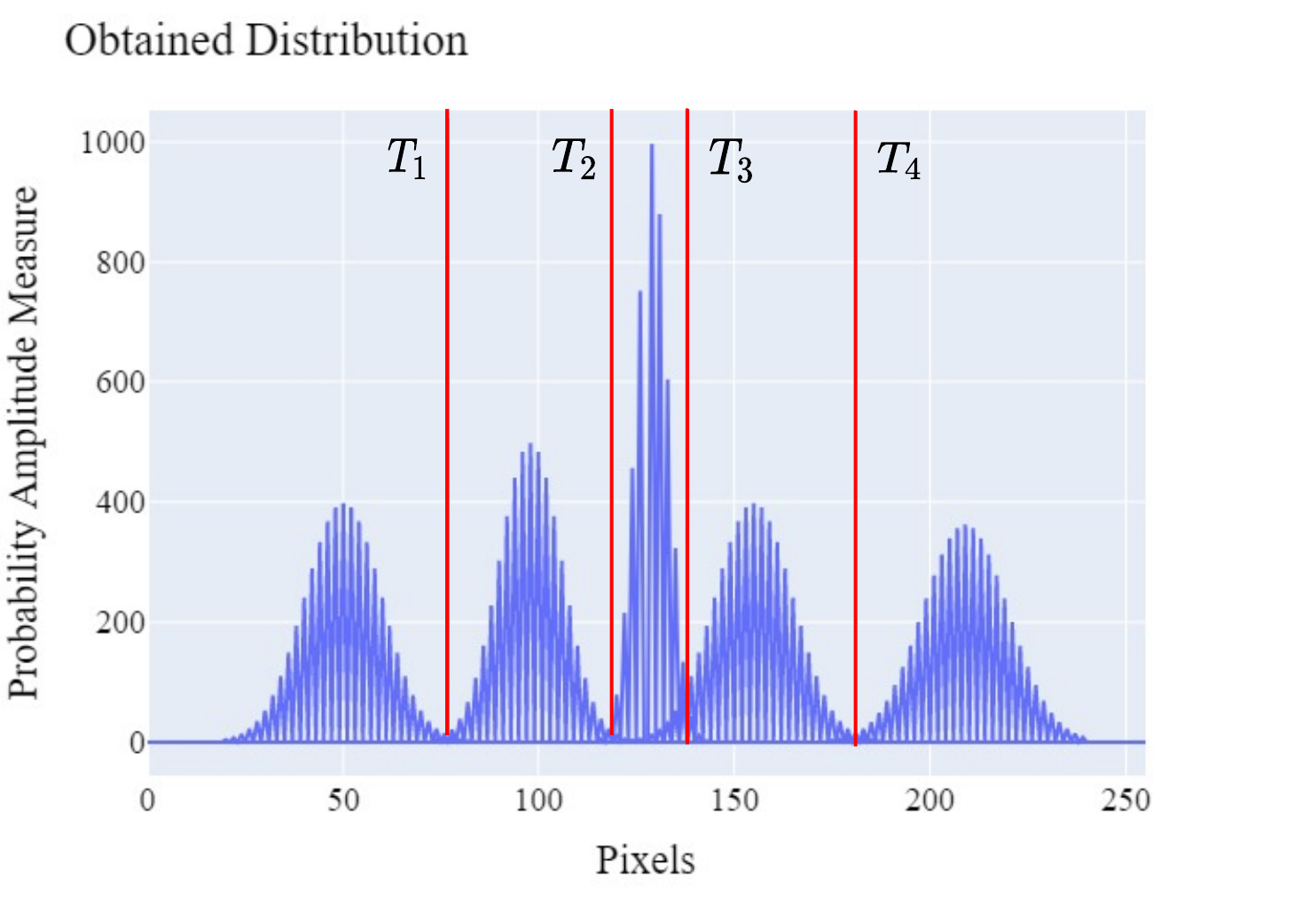} \label{f6b}}
\caption{(a) The histogram of an 8-bit $256 \times 256$ grayscale image of Lena, which contains a much larger range of intensity values, particularly chosen for its pixel distribution, (b) Plot of the measure of probability amplitudes of obtaining different intensity values of their corresponding pixels according to the proposed methodology. Note that the pixel values for which the probability amplitude is 0 are absent in the picture. The obtained thresholds at the intersecting points of the Gaussian are $T_1, T_2, T_3$ and $T_4$. The code implementation of this simulated result can be found on the Github repository given in supplementary material.}
\end{figure}
\textbf{Case-2}\\
A classical Pythonic environment simulates the following results, where the quantum states are taken as vectors, and a matrix formalism is used. In Fig. \ref{f6a}, we have the histogram of Lena, where multiple peaks are seen. According to the given Algorithm \ref{algo1}, $p=5$ for the given histogram and for all the peaks, we select the corresponding values of $i$ and $\delta$ and store them in $r$ and $d$. Now, for each peak, $E_i$ is constructed and operated on $\ket{I}$. This state is simulated and measured, and the results are stored in the dictionary $dict_k$, which stores the probability amplitude as values and the quantum states as keys, as described in the for loop from lines 14 to 20 in Algorithm \ref{algo1}. Note that the number of dictionaries created will equal the number of peaks. So, in this case five quantum circuits are made for five peaks, and their results are stored in five different dictionaries. After getting results for all the peaks, we check the intersecting quantum states, which are common in adjacent dictionaries and have the same probability amplitude. We choose that state as the threshold, realized from lines 14 to 27 of the algorithm. This procedure, in principle, aims to set the thresholds at the valley points of the image histogram for better clarity of the compressed image, highlighting the pixels that are greater in number in the original image. To visually represent the results, we plotted the probability amplitude with the intensity of the five peaks overlapping in Figure \ref{f6b}. The compression quality of different life-like images is discussed in Section \ref{s5}.

\section{Proposed Quantum Circuit for Binarization}
\label{s4}
In this section, we propose a quantum circuit that encodes a $q$-bit $2^n \times 2^n$ digital grayscale image and outputs its binarized form using an improved version of the optimized comparator given in the work of Xia et al. \cite{doi:10.1142/S0217732320500492}. In the mentioned work, the intensity values of pixels are compared with a given threshold using an efficient quantum comparator. Then, the binarized values are represented with their corresponding pixel position values to represent a complete binarized picture. In our work, we intend to decrease the complexity of the circuit by reducing the circuit depth and by using only one qubit to represent the binarized intensity pixel.
\subsection{Image Representation and the Proposed Comparator}
A grayscale image is identified digitally in terms of the position and intensity of all the pixels. If any quantum circuit reproduces these two pieces of information, we get back our original image. So, the qubits are encoded with the position and intensity information in binary strings, i.e., every information is first converted into its binary equivalent and then encoded. Previously, we have seen that the intensity values of an 8-bit grayscale image range from 0-255 pixels of intensity. If we convert 255 to its binary form, we have $2^8$ combinations of 0's and 1's. These different combinations are stored in a function let us call $f_{ij}$ where $f_{ij}$ is represented as $x^0_{ij} x^1_{ij}... x^7_{ij}$, where $x^k_{ij} \in [0,1]$ is either a 0 or a 1 depending on the particular pixel value. It is depicted as:
\begin{equation}
     \ket{f_{ij}} =  \ket{x^0_{ij} x^1_{ij}...x^6_{ij} x^7_{ij}}, 
\end{equation}
This $f_{ij}$ would encode the intensity value from 0 to 255 for the $(i,j)$ pixel. The position information $(i,j)$, where $i$ represents the row number and $j$, the column number, is attached along as tensor products with their corresponding intensity values. The final image representation would be as follows:

\begin{equation}
\label{e16}
    \ket{I}=\frac{1}{2^8} \sum_{i=0}^{2^8-1}  \sum_{j=0}^{2^8-1}\ket{x^0_{ij} x^1_{ij}...x^6_{ij} x^7_{ij}}\ket{ij}
\end{equation}

In Eq. (\ref{e16}), the quantum state $\ket{I}$ serves as a container for all the information, with two distinct components: $\ket{f_{ij}}$ which contains the pixel information in binary format and $\ket{ij}$  holds the position information, also in binary format. This straightforward representation, known as NEQR, allows us to effectively encode all the pixel information onto the qubits, after which they are compared with a given threshold \cite{zhang2013neqr}.\\

\begin{figure}[H]
    \centering
    \includegraphics[scale=0.5]{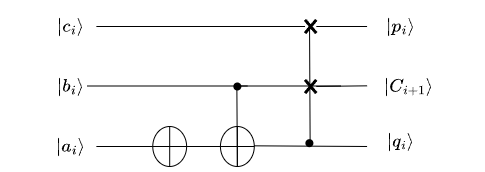}
    \caption{ Circuit diagram of a subtraction slice operation where $a_i$ and $b_i$ are any n-bit strings and $C_{i+1}$ is the output. This idea was proposed by Thomsen et al. \cite{Thomsen_2010}}
    \label{f7}
\end{figure}

Now, a comparator's primary function is to determine the higher value among two provided inputs. This same task could be accomplished using quantum comparators that are connected with the above-prepared image. The simplest quantum comparator for determining the greater of two one-bit inputs is assembled using two Toffoli gates and four qubits (two qubits for the two given inputs and two ancillary qubits). Alternatively, Thomson et al. proposed a version of a quantum comparator in their work, which employs only one ancillary qubit and uses a subtractive borrowing concept \cite{Thomsen_2010}. This modified approach helps determine whether one input value is greater than the other. In Fig. \ref{f7}, we illustrate this configuration, featuring inputs denoted as $a_i$, $b_i$, and $c_i$,  where $a_i$ and $b_i$ each represent $n$-bit strings. The outputs are represented by $p_i$, $C_{i+1}$, and $q_i$. The value of $C_{i+1}$ is referred to as the borrow bit, and any change in its state signifies that $a > b$, while no change implies that $a \leq b$.

\subsection{Implementation}
\label{s4b}
The $n$-bit comparator in Xia's work \cite{doi:10.1142/S0217732320500492} is constructed by recursively calculating the auxiliary bits from the upper levels, combining the idea of ripple carry mentioned in H.-Y. Xia et al.'s work \cite{xia2018efficient}. This technique greatly reduces the consumption of auxiliary bits. We implemented this idea in our work by reducing the depth of the circuit, as our main goal is to determine which of the two input numbers is greater. No inverse operation is done to bring back the outputs to their original state.\\

\begin{figure}[h]
    \centering
    \includegraphics[scale=0.4]{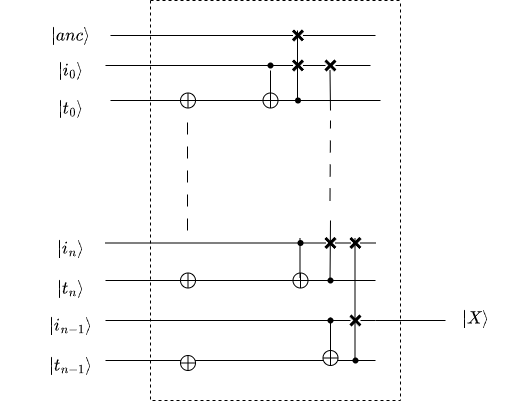}
    \caption{Proposed Quantum Comparator with inputs $\ket{t}$,$\ket{i}$ and output $\ket{x}$. This is the improved version of Xia's comparator \cite{doi:10.1142/S0217732320500492}}
    \label{f8}
\end{figure}

We propose an improved $n$-bit comparator given in Fig. \ref{f8} from Thomsen \cite{Thomsen_2010} and Xia et al's  \cite{doi:10.1142/S0217732320500492} work that takes the binary equivalent of two $n$-bit numbers $\ket{T}=\ket{t_{n-1}...t_1t_{0}}$ and $\ket{I}=\ket{i_{n-1}...i_1i_0}$ and outputs the state $\ket{X}$ which changes if $\ket{I} > \ket{T}$, otherwise it does not change. Given a particular threshold value, it is used to interpret whether the intensity values are higher or lower than the threshold. We use $\ket{X}$ to signify if a particular pixel value $\ket{I}$ is greater than the threshold $\ket{T}$ or not. Extra qubits need not be represented to display binarized states. We attach a $\ket{0}$ or a $\ket{1}$ with the position state instead of a string of $\ket{00...0}$ or $\ket{11...1}$ to represent any of the two intensity values. Encodings are systematically constructed to align with specific requirements and subsequently decoded as needed to extract necessary information. Furthermore, the presented quantum comparator is designed to accept two $n$-bit numbers as input and produce a singular output. The input numbers do not retain their initial state because they become garbage outputs after computation. As a result, we do not increase the circuit depth further to retain the inputs. The primary objective is to determine if an input pixel value is greater than the designated threshold, rendering this approach particularly well-suited to the needs of this work.\\

The computational cost for the whole circuit largely depends on the total number of qubits, which would equal the number of qubits required to encode (all the pixel positions + corresponding intensity values + a threshold value) + one ancillary qubit. For a $q$-bit $2^n \times 2^n$ grayscale image, we would require $2n$ qubits for position encoding, $q$ qubits for intensity encoding, $q$ qubits for threshold encoding, and one qubit for ancillary. So the total number of qubits required equals $[2(n+q)+1]$.

\begin{figure}[H]
\centering
\captionsetup[subfigure]{font=small}
\subfloat[]{\includegraphics[scale=0.35]{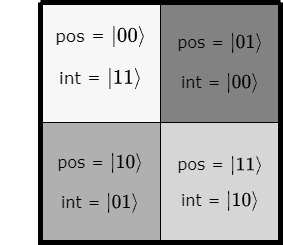} \label{fa}}
\quad
\subfloat[]{\includegraphics[height=0.25\textwidth]{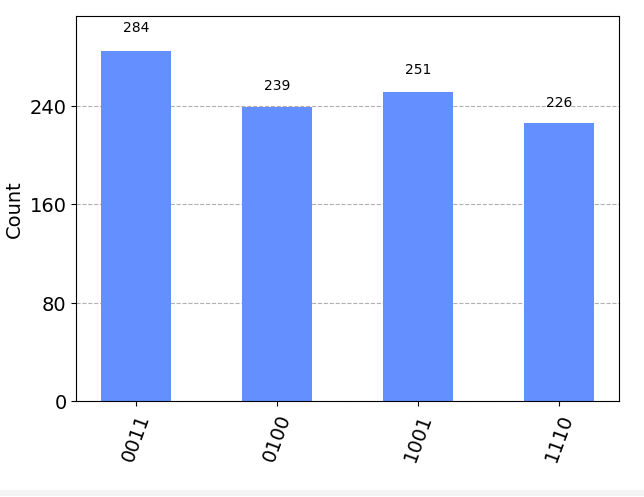} \label{fb}}
\caption{(a) Demo of $2 \times 2$ grayscale image before binarization. (b)This result shows the image represented in the NEQR formalism.  The results show the encoded position by the first two qubits and their corresponding encoded intensity by the last two qubits (the convention is reversed because the qubits are read from left to right in Qiskit) according to the picture given in Fig. \ref{fa}} 
\end{figure}

In the given $2 \times 2$ grayscale image, depicted in Fig. \ref{fa}, we have four pixel values with four types of intensities, as shown in detail in Table \ref{t2}. To define the image, we have $n$=1 and $q$=2. So, the total qubits required for preparing the circuit is 7. We define the position qubits as $\ket{pos_0}$ and $\ket{pos_1}$, the corresponding intensity encoding as $\ket{i_0}$ and $\ket{i_1}$ ,the threshold as $\ket{t_0}$ and $\ket{t_1}$ and finally one ancillary qubit $\ket{anc}$.\\

\begin{table}
    
        \centering
        \caption{Encodings of position and intensity values of the Figure \ref{fa}}
        \begin{tabular}{|c|c|c|c|}
        \hline
        {Space Coordinates} & Encoded String of Intensity value & Intensity value & Overall Reading  \\
        \hline \hline
    {$\ket{00}$} & $\ket{11}$  & 255 & $\ket{0011}$\\
        {$\ket{01}$} & $\ket{00}$ & 0  & $\ket{0100}$\\
        {$\ket{10}$} & $\ket{01}$ & 100 & $\ket{1001}$\\
        {$\ket{11}$} & $\ket{10}$ & 200 &$\ket{1110}$\\
        \hline
       \end{tabular}%
       
       \label{t2}
    
    \hfill
\end{table}
The information from the picture in Fig. \ref{fa} is taken and encoded into a quantum circuit using Qiskit in the NEQR formalism with $\ket{T}$ being the threshold value of binarization and $\ket{I}$ being the image where the state of the image is defined by $\ket{I}=\ket{pos_0pos_1}\ket{i_0i_1}$ given in Eq. (\ref{eq14}) and $\ket{T}=\ket{t_0t_1}$. Table \ref{t2} represents the intensity present in the corresponding pixel position. Here, the exact binary equivalence of the intensity value is not taken as it will increase the number of qubits to encode intensity information. We aim to understand how the qubits are encoded and decoded to interpret the data as a binarized image.

\begin{equation}\label{eq14}
    \ket{I}=\frac{1}{2} (\ket{00}\ket{11} +  \ket{01}\ket{00} +  \ket{10}\ket{01} + \ket{11}\ket{10})
 \end{equation}

\begin{figure*}
    \centering
    \includegraphics[scale=0.19]{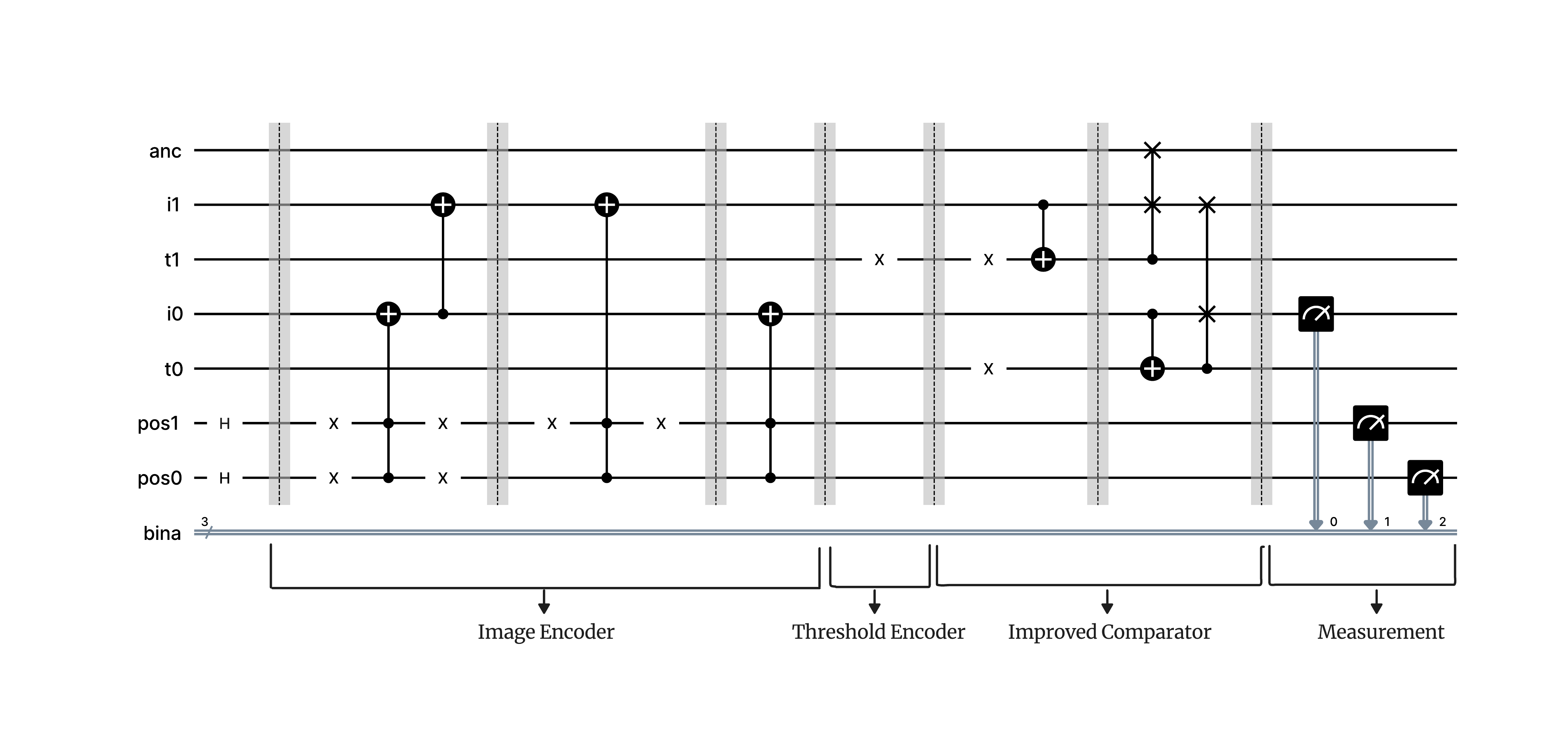}
    \caption{Circuit to binarize the given $2 \times 2$ image. The threshold encoder encodes the threshold value ``01" which is the 100 intensity. The final measurements are done on the output of the comparator $\ket{X}$ and the position qubits $\ket{pos_0}$ and $\ket{pos_1}$}
    \label{f11}
\end{figure*}

\begin{figure}[H]
\centering
\captionsetup[subfigure]{font=small}
\subfloat[]{\includegraphics[height=0.25\textwidth]{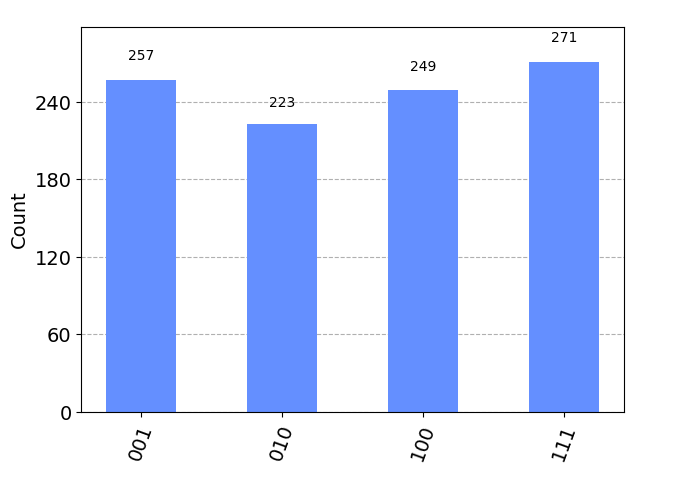} \label{f12a}}
\quad
\subfloat[]{\includegraphics[scale=0.35]{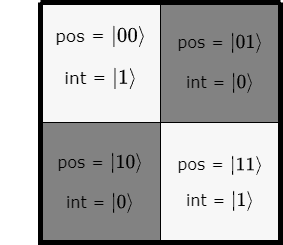} \label{f12b}}
\caption{(a) The results represent the binarized image where the first 2 qubits are the position of the pixel and the last qubit represents the intensity. 0 intensity represents a complete black pixel and 1 represents a complete white pixel. (b) This is the binarized picture we would get from the decoded results of the binarization circuit} 
\end{figure}

The circuit to binarize the constructed image is divided into the following parts, namely the image encoder (pixel position and their corresponding intensities are encoded), the threshold encoder (the threshold value is encoded), the improved comparator (comparator is constructed), and lastly, the measurement, as shown in Fig. \ref{f11}. Fig. \ref{f12a} shows the output, and Fig. \ref{f12b} is the decoded $2 \times 2$ binarized image from the output. The threshold T=$\ket{t_0t_1}$ is set at $\ket{01}=100$, meaning every pixel with a value more than 100 will be binarized to 255 and the rest will be binarized to 0. The output state would either be $\ket{pos_0pos_10}$ or $\ket{pos_0pos_11}$ with the states representing a completely black or a completely white pixel at position  $\ket{pos_0pos_1}$.


In Fig. \ref{f11}, we measure only the output qubit (along with the position qubits), which, in comparison with Fig. \ref{f8}, is the output $\ket{X}$. In contrast, the other qubits are intentionally left unmeasured. It is intuitive that any intensity value exceeding $\ket{01}$ will undergo conversion to the state $\ket{11}$, while values below this threshold will be transformed into $\ket{00}$, which is elaborately discussed in the result.\\

\begin{itemize}
    \item The pixel values in the $\ket{00}$ position were initially at full intensity and remained at full intensity after computation. However, there is a change in the pixel count from $\ket{0011}$ to $\ket{001}$, resulting in a fully illuminated pixel at position $\ket{00}$.

    \item Similarly, the pixel at position $\ket{01}$ retains its complete black state as it transitions from $\ket{0100}$ to $\ket{010}$.

    \item The pixel at position $\ket{10}$ has an intensity of $\ket{01}$ (100), which does not exceed the threshold. Consequently, it is transformed into an entirely black pixel, transitioning from $\ket{1001}$ to $\ket{100}$.

    \item Lastly, the pixel at $\ket{11}$ has an intensity of $\ket{10}$ (200), which surpasses the threshold. Consequently, it transforms into a fully illuminated pixel, transitioning from $\ket{1110}$ to $\ket{111}$.
\end{itemize}
It's worth noting that the intensity space has been reduced to represent a binary image, where pixels are either fully illuminated or completely black. These results are decoded to obtain the final binarized picture, as illustrated in Fig. \ref{f12b}.

\begin{figure}[H]
    \centering
    \includegraphics[scale=0.09]{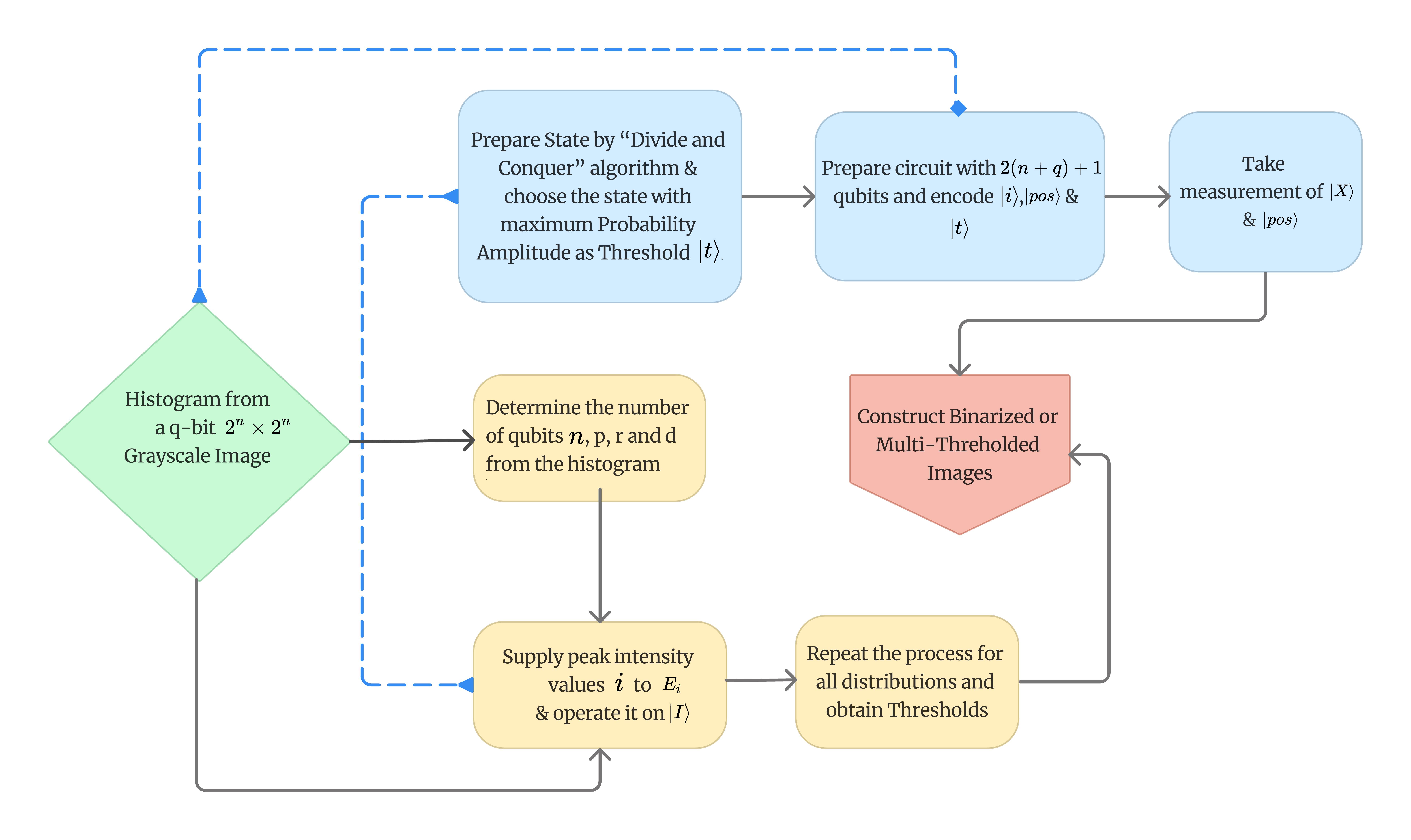}    \caption{Flowchart of the proposed work. The dotted lines refer classical inputs to the quantum circuit }
    \label{f14}
\end{figure}

A flowchart in Fig. \ref{f14} is designed to summarise the whole process, where the hybrid nature of the algorithm is necessary to give vital information to the quantum processor.

\section{Results and Discussions}
\label{s5}
This section discusses the advantages of the proposed methods in this paper, and a flowchart is provided to summarize the multi-thresholding and binarization process comprehensively. The complexity analysis of the constructed circuit is analyzed thoroughly and compared with other works related to this field.
\subsection{Performance Analysis}

\begin{figure}[H]
\centering
        \includegraphics[scale=0.5]{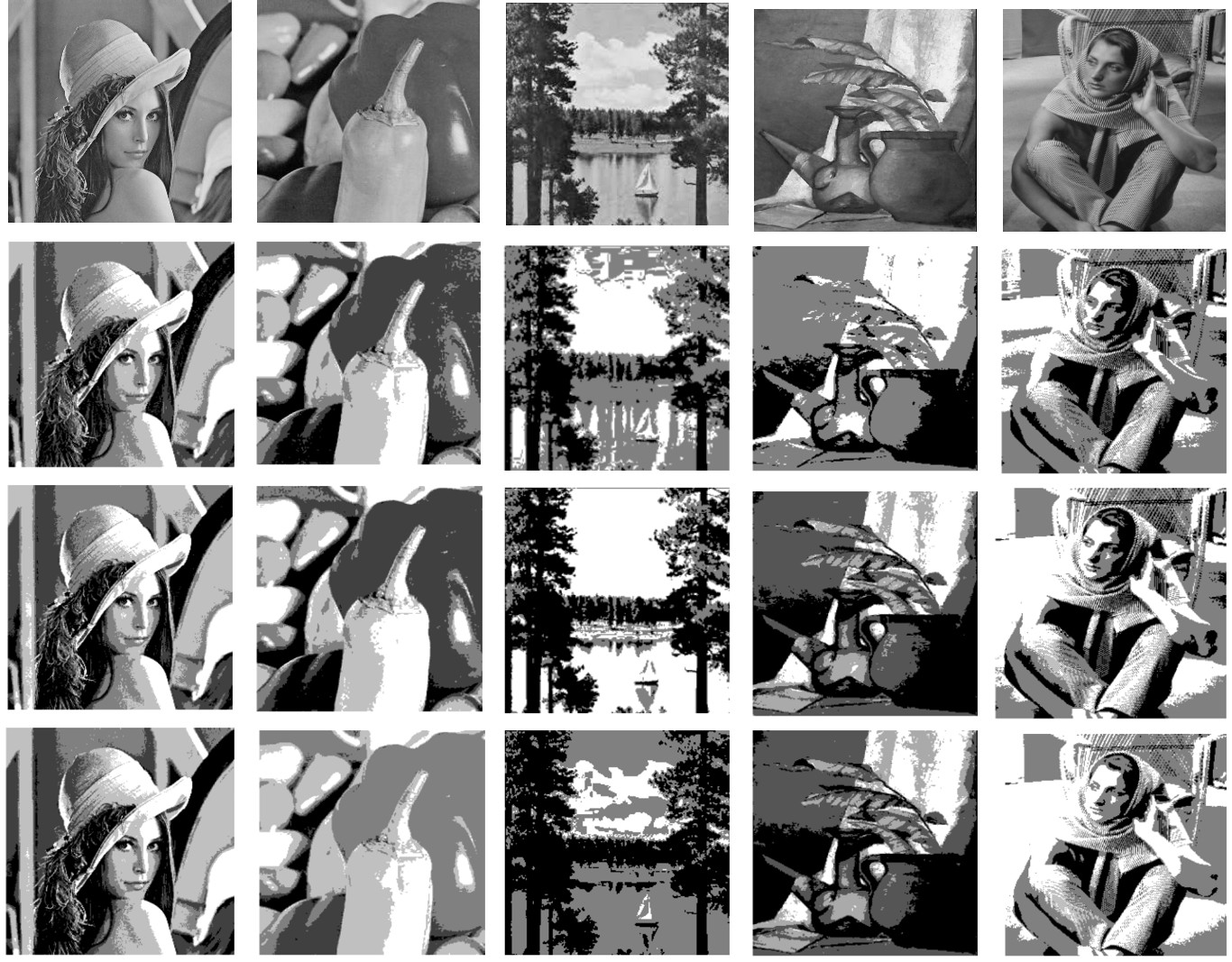}
    \caption{Simulated output of various life-like images. From left, we have the $256 \times 256$ images of Lena, Peppers, Sailboat, Painting and Barbara followed by their respective multi-thresholded images obtained through the recursive algorithm, multi-Otsu algorithm and the proposed methodology respectively in a downward fashion. These images are taken from image processing databases available online.}
    \label{f12}
    
\end{figure}

In Fig. \ref{f12}, we depict the resultant image quality achieved by the proposed multi-thresholding approach on specific intensity values derived from an iterative algorithm applied to the histogram of the provided image. Then we compare the obtained image with images operated on by other algorithms such as the fast Statistical Recursive Algorithm by S. Arora et al. \cite{ARORA2008119} and Otsu's recursive method, also known as multi Otsu algorithm \cite{6313341} in a tabulated form in Table \ref{t3}. The simulations are done in the classical regime, replicating the conditions the algorithm requires as discussed in Section \ref{s3}.

\begin{table}[H]
\begin{center}
    \caption{Comparison Results using different Algorithms}
\label{t3}
\begin{tabular}{|p{2.8cm}|p{3.9cm}|p{3.7cm}|p{3cm}| }
\hline
\multicolumn{4}{|c|}{Lena Comparison Results} \\
\hline
Comparison Metric& Fast Statistical Recursive Algorithm \cite{ARORA2008119}&Multi Otsu Algorithm \cite{6313341}&Proposed Algorithm \\
\hline
PSNR & 28.01 &28.04 & 27.95\\
SSIM &0.2626   & 0.254 & 0.2503\\
\hline
\multicolumn{4}{|c|}{Peppers Comparison Results} \\
\hline
Comparison Metric& Fast Statistical Recursive Algorithm \cite{ARORA2008119}&Multi Otsu Algorithm \cite{6313341}&Proposed Algorithm  \\
\hline
PSNR & 27.8 &27.76 & 27.63\\
SSIM &0.4275   & 0.4568 & 0.5082\\
\hline
\multicolumn{4}{|c|}{Sailboat Comparison Results} \\
\hline
Comparison Metric& Fast Statistical Recursive Algorithm \cite{ARORA2008119}&Multi Otsu Algorithm \cite{6313341}&Proposed Algorithm  \\
\hline
PSNR & 27.628 &27.635 & 27.627\\
SSIM &0.1842   & 0.2322 & 0.1785\\
\hline
\multicolumn{4}{|c|}{Painting Comparison Results}\\
\hline
Comparison Metric& Fast Statistical Recursive Algorithm \cite{ARORA2008119}&Multi Otsu Algorithm \cite{6313341}&Proposed Algorithm  \\
\hline
PSNR & 27.816 &27.813 & 27.739\\
SSIM &0.2762   & 0.2484 & 0.2276\\
\hline
\multicolumn{4}{|c|}{Barbara Comparison Results}\\
\hline
Comparison Metric& Fast Statistical Recursive Algorithm \cite{ARORA2008119}&Multi Otsu Algorithm \cite{6313341}&Proposed Algorithm  \\
\hline
PSNR & 27.778 &27.748 & 27.743\\
SSIM &0.2029   & 0.2210 & 0.2429\\
\hline

\end{tabular}
\end{center}
\end{table}

These pictures are chosen for their spread-out histogram and well-defined peaks, rendering them ideal for histogram analysis. Table \ref{t3} demonstrates the performance of the proposed methodology compared with other well-established algorithms. After applying the different algorithms, the obtained images are rescaled to match the dimensions of their original counterparts and compared with other algorithms. It is evident that the compression quality using UM is at par with other results. In the sailboat picture, the boat is most evident through the UM algorithm than the other ones, while Lena and Barbara can be clearly recognised from the compressed image as well. The results of the similarity index justifies the above statement in Table \ref{t3}. However, there is not a particular algorithm that works best for every image; it depends on what our aim is for processing the image. In Table \ref{t3}, the depicted results of our method do not largely outperform the results of the classical algorithms but are comparable and in some cases perform better. The principal objective of the proposed hybrid-thresholding algorithm is to provide a methodology to extract threshold values by mapping the intensity information onto the probability amplitudes of qubits via applying Gaussian POVM and then computing the threshold value based on the image's histogram. \\ 

\begin{figure}[h]
\centering
        \includegraphics[scale=0.5]{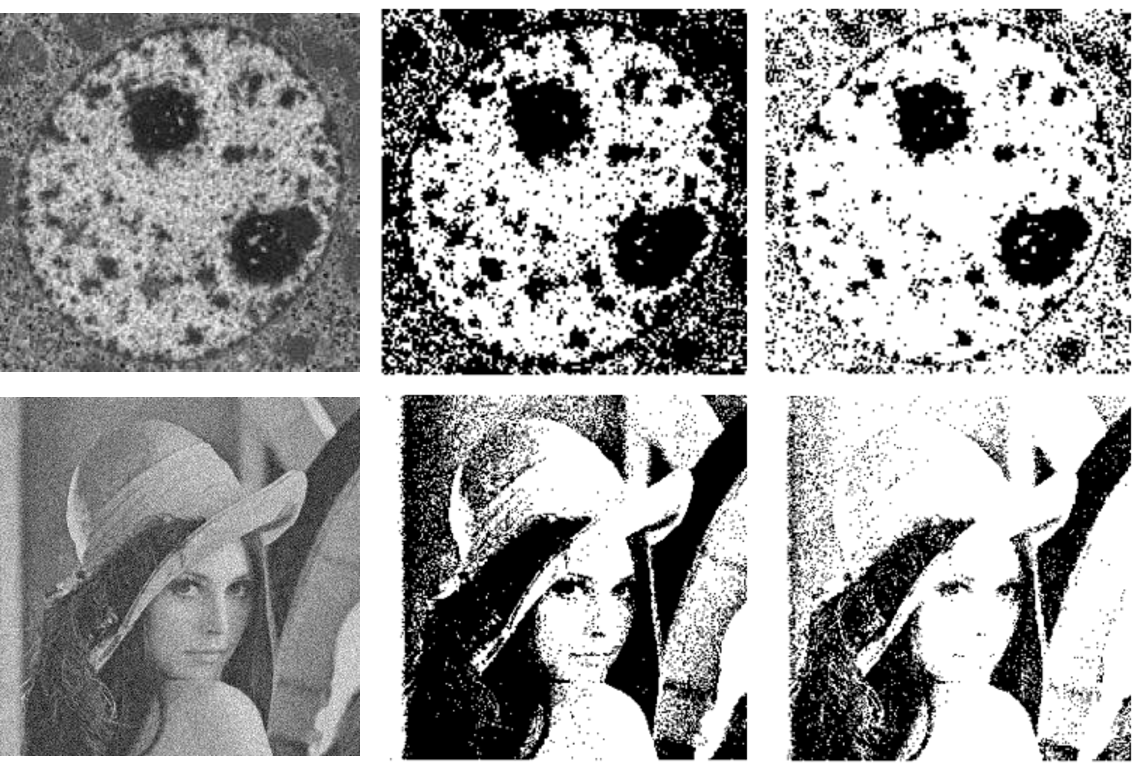}
    \caption{Simulated output of miscellaneous images. We have the $256 \times 256$ images of epoxy resin embedded human tissue with two nucleolus and a noisy image of Lena followed by their respective binarized images obtained through the Otsu algorithm and the proposed methodology respectively starting from left to right.}
    \label{misc}
\end{figure}

In Figure \ref{misc}, more image types have been analyzed, one being a resin embedded human tissue and the other being a noise injected image of Lena. Table \ref{misc} depicts the performance of our algorithm with the Otsu algorithm for binarization \cite{4310076}. The results show that our algorithm performs better with respect to the Otsu algorithm in half of the cases. The tissue cell in the given Figure has two nucleolus, which is prominently identified in the binarized image obtained by the UM algorithm against the background of comparatively less number of black patches. Lena is also identifiable by its binarized form and has a good compression ratio and similarity index compared with the Otsu algorithm. These results show the effective performance of the proposed algorithm even when the histogram of the given image is disturbed, due to its ability to highlight pixels which are greater in number than the other ones, retaining important aspects of the image. 

\begin{table}[H]
\begin{center}
    \caption{Binarization Results using Otsu and the Proposed Algorithm}
\label{bina}
\begin{tabular}{ |p{2.6cm}|p{2cm}|p{2.8cm}|  }
\hline
\multicolumn{3}{|c|}{Human Noise} \\
\hline
Comparison Metric&The Otsu Algorithm \cite{4310076}&Proposed Algorithm \\
\hline
PSNR & 27.789 & 27.813\\
SSIM &0.055 & 0.0514\\
\hline
\multicolumn{3}{|c|}{Lena Noise} \\
\hline
Comparison Metric&The Otsu Algorithm \cite{4310076}&Proposed Algorithm  \\
\hline
PSNR & 27.773& 27.899\\
SSIM &0.0716 & 0.0936\\
\hline

\end{tabular}
\end{center}
\end{table}


\subsection{Complexity Analysis}
From the procedures involved in Algorithm \ref{algo1} we indicate that the complexity is of the order of $O(p2^{2q})$ for a $q$-bit $2^n \times 2^n$ image where $p$ is the number of peaks present in the image histogram. This is evident from the nested for loops present in lines 21-27 in Algorithm \ref{algo1}. The divide and conquer load circuit has complexity of order $O(q^2)$ as discussed in the reference of Aruajo et al \cite{araujo2021divide}.

The proposed quantum circuit utilizes the NEQR framework to prepare images. The preparation cost for generating a $q$-bit $2^n \times 2^n$ image within this framework is $2n+q$, where $q$ represents the number of color bits used. The time complexity involved to prepare the quantum image in NEQR formalism is $O(qn2^{2n})$. To prepare the comparator, $q$ qubits are needed to encode the threshold and one ancillary qubit. Consequently, the total number of required qubits is $2(n+q)+1$, as discussed in section \ref{s4b}. To create the optimized quantum comparator, $q$ control switch gates, $q$ CNOT gates, and $q$ NOT gates are required. The quantum cost for a single NOT gate and CNOT gate is 1, while the quantum cost for a control switch gate is five \cite{doi:10.1142/S0217732320500492}. Therefore, the overall cost amounts to 7$q$, excluding the costs associated with quantum image preparation, encoding the threshold, and the measurement operations. This optimized comparator does not restore the inputs to their original state, which aligns with the specific requirements of this work. Comparative analyses with other quantum comparators are shown in Table \ref{t4}.\\

\begin{table}[H]
\caption{Comparison Results using different algorithms}
\label{t4}

\begin{tabular}{ p{3.2cm} p{2.8cm} p{2.8cm} p{2.8cm}}
\hline
Algorithm & Quantum Cost &Quantum Delay&Auxiliary bits \\
\hline
Wang et al.\cite{wang2012design} & $O(n^2)$ &$O(n^2)$ & $2n-2$\\
Al-Rabadi \cite{al2009closed}& $39n+9$ &$24n+9$  & $6n+1$ \\
Thapliyal et al. \cite{5697872} &$18n+1$   & $18 \log _2{n} + 7$ & $4n-3$\\
Vudadha et al \cite{6296477}& $14n$   & $5 \log_2n + 12$ & $4n-2$\\
Xia et al. \cite{doi:10.1142/S0217732320500492}& $13n + 1$   & $2(n+1)+3$ & 1 (corrected with $2n+2$)\\
Shiping Du et al. \cite{DU2022105710}& $9n-1$   & $2n+7$ & $2n+1$\\
Proposed Comparator& $7n$   & $n+2$ & 1\\
\hline
\end{tabular}

\end{table}

The reduced complexity of the proposed Quantum comparator is because it does not change the inputs back to their original state; it is only able to determine whether $\ket{i}>\ket{t}$  and nothing else. Also, it does not represent the original $n$-string input; instead, it uses only 1 bit to represent the intensity, which matches our aim to reduce the circuit depth.\\

In summary the total circuit complexity involved in determining the thresholds, encoding and then segmenting a $q$-bit $2^n \times 2^n$ with $p$ peaks present in its image histogram is given by Eq.(\ref{complexity}).
\begin{equation}
\label{complexity}
    O(p2^{2q}) + O(q^2) + O(qn2^{2n}) + (n+2)
\end{equation}

This complexity result is better than most of the classical image processing algorithms present today.

\section{Conclusion}
\label{s6}
In conclusion, we have used unsharp measurements as a tool to interpret normal distributions of different widths that are naturally evident in an image histogram. Gaussian POVMs are constructed and operated on intended intensity values encoded in quantum states, and using this we construct the effect operators and a quantum circuit by the ``Divide and Conquer" algorithm. It is measured to reveal the probability amplitude of quantum states. For an unimodal histogram, the state with the highest probability amplitude is chosen as the threshold; otherwise, the thresholds are set as the states which are common and have the same count.\\

The number of thresholds is inherently determined by the image's histogram, giving this procedure an advantage over many other algorithms. The proposed algorithm is applied to various life-like and noisy images. The results demonstrate sharp contrast in critical image features, with compression and similarity ratios equivalent to established classical algorithms, as measured by metrics such as PSNR and SSIM while retaining important characteristics of the original image. An improved quantum comparator is demonstrated, which compares the pixel value with a threshold and generates only one state as output. The binary image is represented by measuring only the position pixel qubits and a single measurement of the output of the quantum comparator; as a result, the quantum cost and quantum delay of the quantum comparator are significantly lower than any of its previous counterparts. It would be interesting to investigate further the integration of unsupervised learning techniques to predict the width and mean of the constructed Gaussians depending on a given image histogram and to construct gate-based POVM operators for compact circuit implementation and, hence, a possible design of a more robust algorithm that can perform more efficiently on present-day noisy quantum computers.\newline 
\\
{\small \textbf{Acknowledgements} \hspace{0.06cm}  AB would like to thank Prasanta K. Panigrahi and IISER Kolkata for their warm hospitality and for providing various resources essential for this work. The authors would like to thank Divyanshi Dwivedi for her valuable contributions. PKP acknowledges the support from DST, India, through Grant No. DST/ICPS/ QuST/Theme-1/2019/2020-21/01.} 

\section*{Data Availability}
 The site \href{https://github.com/AyanBarui/A-Novel-Approach-to-Threshold-Images-using-Unsharp-Measurements/tree/main}{Github} contains the codes used in this study.

\section*{Declarations}
{\small \textbf{Conflict of interest} \hspace{0.06cm} We declare that we have no known competing interests or personal relationships that could have appeared to influence the work reported in this paper.}

\bibliographystyle{elsarticle-num-names}
\bibliography{Revised}

\end{document}